\documentclass[fleqn,10pt,twocolumn]{article}
\usepackage[toc,page]{appendix}
\usepackage{authblk}
\usepackage{amsmath}
\usepackage{abstract}
\usepackage{overpic}
\usepackage{booktabs}
\usepackage{multirow}
\usepackage{cleveref}
\usepackage[usenames,dvipsnames]{xcolor}
\usepackage{xspace}
\usepackage{tikz}

\newcommand{\et}{\textit{et al.}\xspace}
\newcommand{\ie}{\textit{i.e.}\space}
\newcommand{\eg}{\textit{e.g.}}

\title{The Effect of Recency to Human Mobility}

\author[1,*]{Hugo Barbosa}
\author[2]{Fernando B. de Lima-Neto}
\author[3]{Alexandre Evsukoff}
\author[1]{Ronaldo Menezes}
\affil[1]{BioComplex Lab, Department of Computer Sciences, Florida Institute of Technology,  USA}
\affil[2]{Computational Intelligence Research Group, Polytechnic School, University of Pernambuco,   Brazil}
\affil[3]{COPPE, Federal University of Rio de Janeiro, Brazil}

\affil[*]{hbarbosa@biocomplexlab.org}

\date{}
\begin{document}
\twocolumn[
\maketitle
\begin{onecolabstract}
In recent years, we have seen scientists attempt to model and explain human dynamics and in particular human movement. Many aspects of our complex life are affected by human movement such as disease spread and epidemics modeling, city planning, wireless network development, and disaster relief, to name a few. Given the myriad of applications it is clear that a complete understanding of how people move in space can lead to huge benefits to our society. In most of the recent works, scientists have focused on the idea that people movements are biased towards frequently-visited locations. According to them, human movement is based on a exploration/exploitation dichotomy in which individuals choose new locations (exploration) or return to frequently-visited locations (exploitation). In this work we focus on the concept of {\it recency}. We propose a model in which exploitation in human movement also considers recently-visited locations and not solely frequently-visited locations. We test our hypothesis against different empirical data of human mobility and show that our proposed model is able to better explain the human trajectories in these datasets.\\


\end{onecolabstract}
]



\section*{Introduction}
The understanding on the fundamental mechanisms governing human mobility is of importance for many research fields such as epidemic modeling \cite{Belik2011,Colizza2007,Balcan2011}, urban planning\cite{Toole2012,Lenormand2015}, and traffic engineering \cite{Kitamura2000,Jung2008,Krajzewicz2011a}. Although individual human trajectories can seem unpredictable and intricate for an external observer, human trajectories are , in fact, very predictable \cite{Song2010b,Wang2011,Yang2014,Sadilek2012,Krumme2013,Lu2013} and regular over space and time \cite{Gonzalez2008a,Brockmann2006,Hasan2012a}. One characteristic of human motion, largely observed in empirical data, is the fact that we have the tendency to spend most of our time in just a few locations \cite{Gonzalez2008a,Song2010,Schneider2013}. More generally, the distribution of visitations frequencies have been observed to be heavy tailed \cite{Song2010,Krumme2013}. 

However, the fundamental mechanisms responsible for shaping our visitation preferences are still not fully understood. The \emph{preferential return} (PR) mechanism, proposed by Song \et \cite{Song2010}, offered an elegant and robust model for the visitation frequency distribution.  It defines the probability $\Pi_{i}$ for returning to a location $i$ as $\Pi_{i}\propto f_{i}$, where $f_{i}$ is the visitation frequency of the location $i$. It implies that the more visits a location receives, the more visits it is going to receive in the future, which in different fields goes by the names of \emph{Matthew effect} \cite{Merton1968}, \emph{cumulative advantage} \cite{Price1976}, or \emph{preferential attachment} \cite{Barabasi1999}. 

Although the focus of the PR mechanism---as part of the Exploration and Preferential Return (EPR) individual mobility model---was used to reproduce some of the scaling properties of human mobility, its general principles are grounded on implausible assumptions from the human behavior point of view. In the long term, the PR assumption as a property of human motion leads to two discrepancies. First, in the model, the earlier a location is discovered, the more visits it is going to receive. In implies that first visited location will most likely also be the most visited one. Second, if the cumulative advantage indeed holds true for human movements, people would not change their preferences, which is clearly not true.  

Here we propose that {\it the PR mechanism has to simultaneously consider the frequencies of visits and the time of these visitations in individual human trajectories}. Using mobility data obtained from call detail records (CDRs) and location-based check-ins produced by thousands of users, we uncover a strong tendency of individuals to return to recently-visited locations, a behavior similar to the one observed by Szell \et in a virtual world \cite{Szell2012}. Moreover, we show that such tendency is not conditioned to the previous visitation frequencies. Last, we introduce a variation to the EPR model to incorporate the influence of recency in individual trajectories. Our approach is based on the empirical evidences that the longer the time since the last visit to a location, the lower is the probability of observing a user at this location \cite{Song2010,Gonzalez2008a}.  

\section*{Materials and Methods}
\subsection*{Data}

In this work, we used two mobility datasets: the first one ($D1$)  corresponds to 6 months of anonymized mobile-phone traces from a large metropolitan area in Brazil. This dataset is composed of 8,898,108 records from 30,000 users between January 1--June 30, 2014. The second dataset ($D2$) is composed of 23,736,435 \emph{check-ins} from 51,406 Brightkite users in 772,966 different locations.\footnote{Brightkite was a location-based social networking service launched in 2007 and closed in 2011 \cite{Grabowicz2013,Cho2011}.} Unlike the mobile phone data, locations in the Brightkite dataset correspond to the actual places where the users checked in---phone data locations correspond to the antena tower the phone communicates with and hence are approximations of the user's actual location.

Since our interest here is on the individuals' \emph{trajectories}, in this analysis we considered only the data that provides information relating to the users' displacement. Hence, we filtered out repeated observations in one place, resulting in a time series for each individual representing their trajectories over the observed period. For instance, if we assume $A$, $B$ and $C$ are locations and the data shows the a user in the locations (in this order) $[A,B,B,B,C,C,A,A,A,B]$ the trajectory is considered to be $[A,B,C,A,B]$ because users remaining in the same location between consecutive data points are not considered to have ``moved''.
Furthermore, to reduce the influence of co-located antennas (common in densely-populated sites), we merged those within less than 10 meters apart under the just one id.

\subsubsection*{Heterogeneities in human mobility}

The first analysis we performed measures the population-level heterogeneities represented by the different activity patterns. First we determine the number of  observed displacements ($N$) per user during the period considered. Notice that it does not necessarily represent the actual number of displacements, but rather the number of \emph{jumps} per user captured by the datasets. All the scaling parameters were estimated using the methods described by Clauset \et \cite{Clauset2009}.

\begin{figure}[htbp]
 \centering
 \includegraphics[width=0.9\columnwidth]{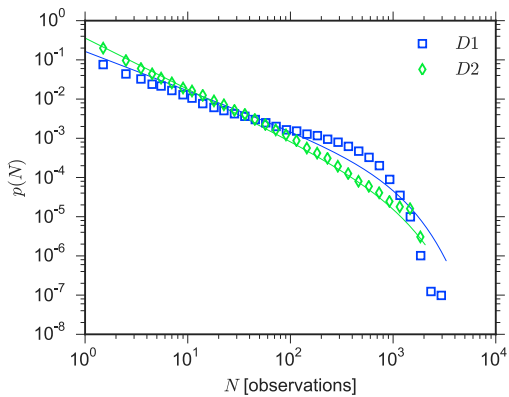}
 \caption{ \textbf{The Number of observed displacements per user.}  The probability density function of the number of observed displacements during the observational period.}
\label{fig:n_of_visits}
\end{figure}

The $p(N)$ of $D1$ and $D2$ are better approximated by truncated power-law distributions, defined as $p(x) = Cx^{-\alpha}\mathrm{e}^{-x/\tau}$ whose parameters where estimated using the maximum likelihood method (see \cref{appendix:stats} for statistical validation). For $D1$, the exponents were found to be $\alpha_{D1}\approx 1.0$ and $\tau_{D1} \approx 783$ observations whereas for $D2$ the exponent are $\alpha_{D2} \approx 1.3 $ and $\tau_{D2} \approx 923$ observations (see Figure~\ref{fig:n_of_visits}). This means that in both datasets, users tend to not move a lot, and highly mobile individuals are very rare. For instance, in $D1$, the daily average number of displacements is approximately 2.2 whereas in $D2$ it was approximately 1.7. The average number of jumps per month in $D1$ is 24.5 while in $D2$ it was 9.2. The lower average number of movements in $D2$ could be because Brighkite was a location-based social networking service, hence, movements related to social activities must be overrepresented in it. Nevertheless, given our focus is on individuals' visitation \emph{preferences}---rather than \emph{needs}---it does not affect negatively our analysis.

\begin{figure}[htbp]
  \centering \includegraphics[width=0.9\columnwidth]{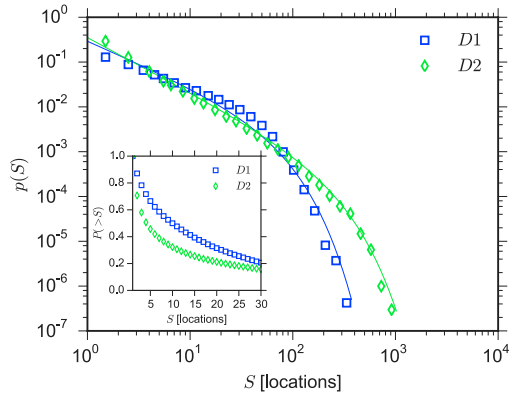}
  \caption{\textbf{Number of distinct visited locations}. The probability density function of the number of unique visited locations aggregated by users. Solid lines correspond to the best fits. The inset is the CCDF of the distribution in linear scale, illustrating the fact that people tend to concentrate most of their visits to just a few locations.} 
  \label{fig:n_locations}
\end{figure}

From the human mobility perspective, we extracted the number of {\it distinct} locations users have visited in the period (Figure \ref{fig:n_locations}).  It depicts the probability $p(S)$ of a user having visited $S$ distinct locations at the end of the observational period.  For $D1$, the number of visited locations is better fitted by a log-normal distribution with parameters $\mu_{D1}\approx 3.16$ and $\sigma_{D1} \approx 0.73 $  while $D2$ follows a truncated power law whose exponents are $\alpha_{D2} \approx 1.22$ and $\tau_{D2} \approx 200.0$. When we look at the CCDF (Complementary Cumulative Distribution Function) in linear scale (inset of Figure \ref{fig:n_locations}) it becomes even more evident the fact that we spend most of our time in a very few locations. To illustrate, about 30 \% of the time, users in $D1$ were found at just 2 locations while in $D2$ this number was approximately 40\%.

\subsubsection*{Temporal patterns}
\begin{figure}[!htbp]
\centering
\includegraphics[width=\columnwidth]{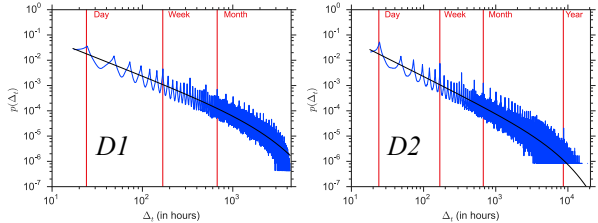}
\caption{ \textbf{Return probabilities as a function of the elapsed time $\Delta_{t}$ since the most recent visit.} Peaks are observed at 24h intervals, capturing the temporal regularity of which humans return to previously visited locations. Also, it is possible to see that the return probability decays very quickly as the time increases. Solid lines correspond to the truncated power-law fits with exponents $\beta_{D1} \approx 1.405$ and $\beta_{D2} \approx 1.425$.   }
\label{fig:delta_t}
\end{figure}

In a modern society, where most of the people have daily routines, part of our trajectories are constrained to a limited number of locations at regular time intervals. Human activity routines are responsible for part of the regularities manifested in human movements. From the empirical data, we extracted the time interval (in hours) between two consecutive visits to a location. The distribution of time intervals is depicted in Figure \ref{fig:delta_t}. The plot reveals two important features of human movements: first, one can observe existence of peaks in 24h intervals representing the users' daily routines \cite{Gonzalez2008a}. Additionally, we can see the presence of weekly  repetitive patterns as previously observed \cite{Hsu2006}. More formally, the probability of returning to a location decreases with $p(\Delta_{t}) \propto \Delta_{t}^{-\beta}\mathrm{e}^{-\Delta_{t}/\kappa}$ with  $\beta_{D1} \approx 1.405$ and $\kappa_{D1} \approx 2,189$ hours and $\beta_{D2} \approx 1.425$ with $\kappa_{D2} \approx 6,791$ hours. The second one is the fact that both distributions exhibited very similar power-law exponents $\beta$, even though the two datasets are very different in terms of coverage, spatial resolution,  acquisition method and time span. It suggests that the temporal dimension of return mobility movements are scale invariant, supporting the general nature of our findings.

\section*{Results}
\subsection*{A rank-based analysis of human visitation patterns}

In this work, we propose a rank-based approach to the analysis of human trajectories. For such, we defined two rank variables, namely the \emph{frequency} rank ($K_{f}$) and the \emph{recency} rank ($K_{s}$). Both ranks were measured in a expanding basis from the accumulated sub-trajectories. To illustrate, consider a particular user $x$ with a trajectory $T=[(l_{1},l_{2},\ldots,l_{n}),l_{i}\in[1,\ldots,N]]$ composed of $n$ steps to $S \le N$ locations. For each step $j>0$, we have the partial trajectory $\mathcal{T} = [l_{1},l_{2},\ldots, l_{j-1}]$ composed of all the previous steps, with $l_{j-1}$ being the immediate preceding step. From the sub-trajectory $\mathcal{T}$ we compute the frequency-based ranks $K_{f}$ of all locations visited so far. If the step $l_{j}$ is a return (\ie, $l_{j}\in \mathcal{T}$) we say that the frequency rank  of the location $l_{j}$ is the rank $k_{f}(j) = K_{f}[l_{j}]$. 

As previously described, the PR mechanism suggests that the visitation probability of a particular location is proportional to the number of previous visits to it ($K_{f}$). Our claim is that the Zipf's Law observed in visitation frequencies distribution is influenced by our \emph{tendency to return to recently visited locations} ($K_{s}$).

 To test such influence we compared the return probabilities from two ranking approaches: one based on the visitation frequencies ($K_{f}$) and the other based on the recency of the last visit to a location ($K_{s}$).  
 
 In summary, the two ranks can be described as:
 
 \begin{itemize}
 \item $\mathbf{K_{s}}$ is the recency-based rank. A location with $K_{s} = 1$ at time $t$ means that it was the \textbf{previous} visited location. $K_{s} = 2$ means that such location was the second-most-recent location visited up to time $t$ and so on.
 
 \item $\mathbf{K_{f}}$ is the frequency-based rank. A location with $K_{f} = 1$ at time $t$ means that it was the \textbf{most} visited location up to that point in time. Similarly, a location with $K_{f} = 2$ is the second-most-visited location up to time $t$.
 \end{itemize}

\begin{figure*}[htbp]
 \centering
\includegraphics[width=0.9\linewidth]{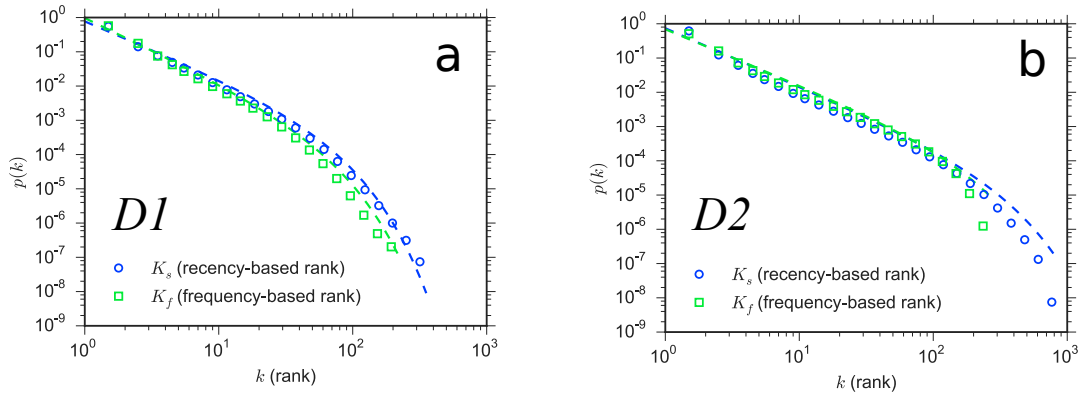}
 \caption{\textbf{Comparison between the probability of return by recency and frequency ranks.} The distributions of both ranks can be better approximated by truncated power laws (dashed lines).  \textbf{(a)}  The recency-based rank of $D1$	has  exponents  $\alpha_{K_{s}} \approx 1.644$ and exponential cut-off $\tau_{K_{s}}\approx 41.66$ , whereas the frequency-based rank distribution has a better fit for $\alpha_{K_{f}}\approx 1.86$  with $\tau_{K_{f}} \approx 37$. \textbf{(b)} The best fit for the return ranks distribution in $D2$ is achieved with parameters $\alpha_{K_{s}} \approx 1.699$ and  $\tau_{K_{s}}\approx 250$ for the recency rank whereas the frequency rank has parameters  $\alpha_{K_{f}}\approx 1.625$  and $\tau_{K_{f}} \approx 125$}
 \label{fig:order_order_min_rank_d1_order_min_rank}
\end{figure*}

Given the definitions above, we first analyzed the probability of return as a function of $K_{s}$. This analysis shows that such probability  decays vary rapidly with $K_{s}$ (Figure \ref{fig:order_order_min_rank_d1_order_min_rank}). More precisely, for $D1$, the probability $p(K_{s})$  follows a truncated power law,  with exponent $\alpha_{K_{s}} \approx 1.644$  whereas the best fit for the frequency-based rank distribution is achieved when $\alpha_{K_{f}}\approx 1.86$. For $D2$, the best fit for the return ranks distribution in $D2$ is obtained with parameters $\alpha_{K_{s}} \approx 1.699$ for the recency rank, whereas the frequency rank has the exponent $\alpha_{K_{f}}\approx 1.625$ (see \cref{appendix:parameters} for details on the curve fitting methods and results). Similarly to what was observed on the distributions of inter-return times $p(\Delta_{t})$, the scaling exponents observed in the recency rank distribution are very similar. It suggests that the recency rank may capture a fundamental mechanism underlying the return movements.



\subsection*{Recency over frequency: the role of recent events in human mobility}
In this section we explore the two-dimensional density distribution of returns  $p(K_{f},K_{s})$. The idea is to investigate the return probabilities as an outcome of the convolution between visitation frequencies and times, encoded in $K_{f}$ and $K_{s}$ simultaneously. 
If users have a stronger preference for recently visited locations we should observe:
\begin{enumerate}
\item lower values of $K_{s}$ must be frequently observed over a wider range of $K_{f}$. It would suggest that we tend to return to recently visited locations even if we have not visited such location many times before (\ie, lower $K_{f}$ rank);
\item higher values of $K_{f}$ must deviate from lower $K_{f}$ values, suggesting that the probability of return to a location decays with time, even if it was a highly visited location.
\end{enumerate} 

\begin{figure*}[htb]
\centering
\includegraphics[width=\textwidth]{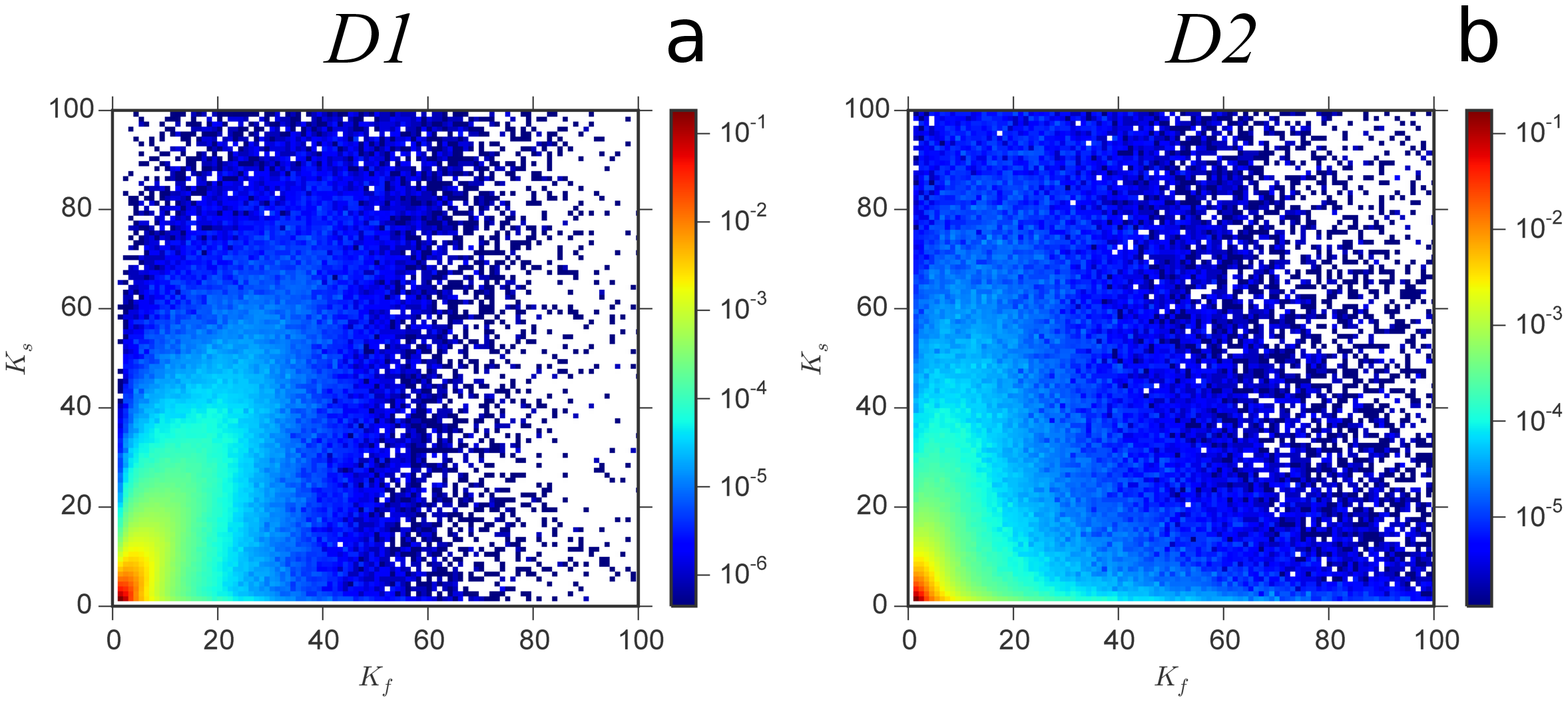}\\
\includegraphics[width=.8\textwidth]{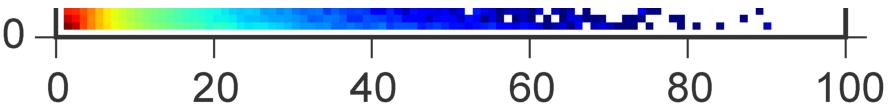}\\
\includegraphics[width=.8\textwidth]{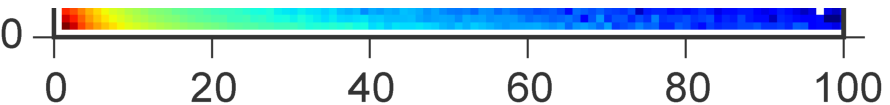}
\caption{\textbf{Return probabilities}. Each point represents a return, whereas the color encodes the density of points. The ranks here were shifted to have the highest-ranked locations at $(0,0)$.  A point $(x,y)$ in the histogram represents a return to the $(x+1)^{\text{th}}$ most visited location after $y+1$ steps. \textbf{(a)} Looking at the return ranks distribution for $D1$ we can observe that the recency influence is less pronounced in $D1$ in comparison with $D2$. \textbf{(b)}  On the other hand, the fine-grained data of $D2$ shows a strong influence of recency. }
\label{fig:heatmaps}
\end{figure*}

\begin{figure*}[htb]
\centering
\includegraphics[width=0.9\textwidth]{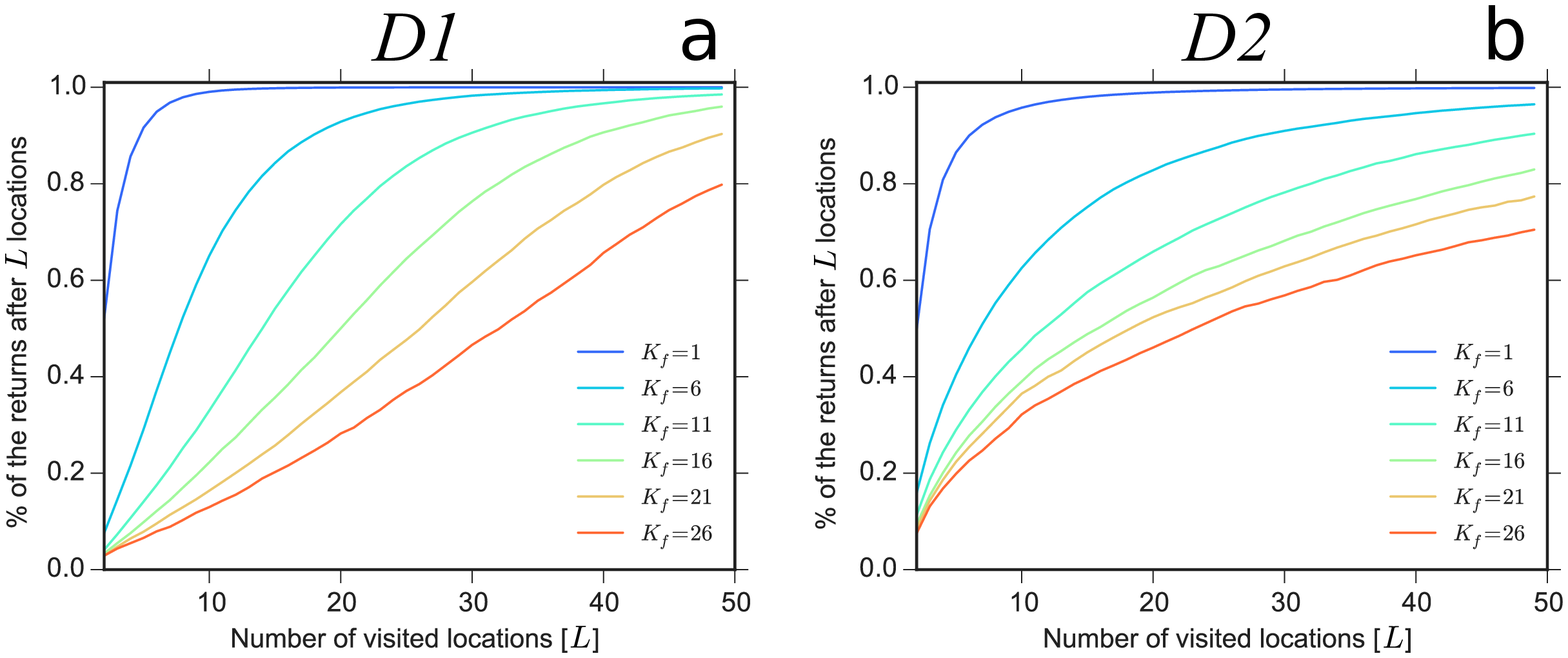}
\caption{\textbf{Fraction of returns to the $K_{f}$ most visited location occurring after the visitation of $L$ different locations.}  Another way to look at the recency effect is by analyzing the correlation between the number of different visited locations between two visits to a location. We can see that people tend to return to their most visited locations after visiting very few places. \textbf{(a)} In $D1$, more than 86\% of the returns to the most visited location occurred after visiting less than five other locations while for $D2$ \textbf{(b)} it was more than 91\%.}
\label{fig:return_ratios}
\end{figure*}

To test these hypotheses, we analyzed the frequency of returns with ranks $(K_{f}, K_{s})$ for all $K_{f}$ and $K_{s}$ values. For example, a visit to a location with ranks $(10,3)$ means a return to the 10$^\text{th}$ most visited site after visiting $3$ other locations. This return distribution is represented as a two-dimensional histogram (shown as heatmaps) for each of the datasets (Fig. \ref{fig:heatmaps}). From the heatmaps, we can observe that returns to the most visited locations (\eg, $K_{f} \le 7$) have shorter return trajectories. In other words, when it comes to our most visited locations, we tend to return to them after visiting very few locations. It can be seen by the rapid decrease in the returns frequencies when $K_{s}$ grows. For instance,  in $D1$, more than 86\% of the returns to the most visited location occurred after visiting less than five other locations while for $D2$, it was more than 91\% (see Figure \ref{fig:return_ratios}).

We can observe also that the recency increases the probability of return to less visited locations (\eg, $7\le K_{f}\le 40$), expressed by a broader distribution of $K_{f}$ when $K_{s}$ is low (\eg, $K_{s} \le 3$). For instance, a closer look at the bottom rows of the plots in Figure \ref{fig:heatmaps} (in the detail) shows that a recent visit to a location can increase the probability of returning to it up to 10 times  in $D2$ (see Figure \ref{fig:heatmaps}b).

When we compare $D1$ and $D2$  we can observe a slightly different pattern between them. First, the effect of recency is much stronger in $D2$ than in $D1$. Such difference could be explained by fact that the mobility data of $D1$ is coarse-grained to a cell-tower level. $D2$, on the other hand, provides finer-grained mobility data, capturing changes in visitation preferences, independent on the distance between the locations.

Additionally, in order to verify whether the power law observed in the recency rank distribution is rooted on the temporal semantics of individuals' trajectories, we applied our rank-based approach to randomized versions of both empirical datasets ($D1$ and $D2$). The first randomized dataset we analyzed ($R1$) was obtained from uniformly shuffling each individual trajectory. This way, we artificially remove any temporal information possibly encoded within the individual trajectories, while maintaining the visitation frequencies intact. On the second randomization method ($R2$), we also remove the visitation frequencies by generating for each user a new random trajectory with the same number of displacements, and the same number of distinct visited locations. To serve as the baseline for the analyses, the data of the third randomization approach ($R3$) produces a new dataset with the same size as the original one, but keeping only the total number of users and locations. More precisely, for each of the datasets, we generated a randomized version of them with $M$ random points $$v_{m} = [u_{m},l_{m},m],m\in[1,\dots,M], $$ where each $u_{m}$, $l_{m}$ is uniformly sampled from $U$ users and $N$ locations respectively, with $M$, $U$ and $L$ the same as in $D1$ and $D2$.

\begin{figure*}[ht]
 \centering
\includegraphics[width=0.9\linewidth]{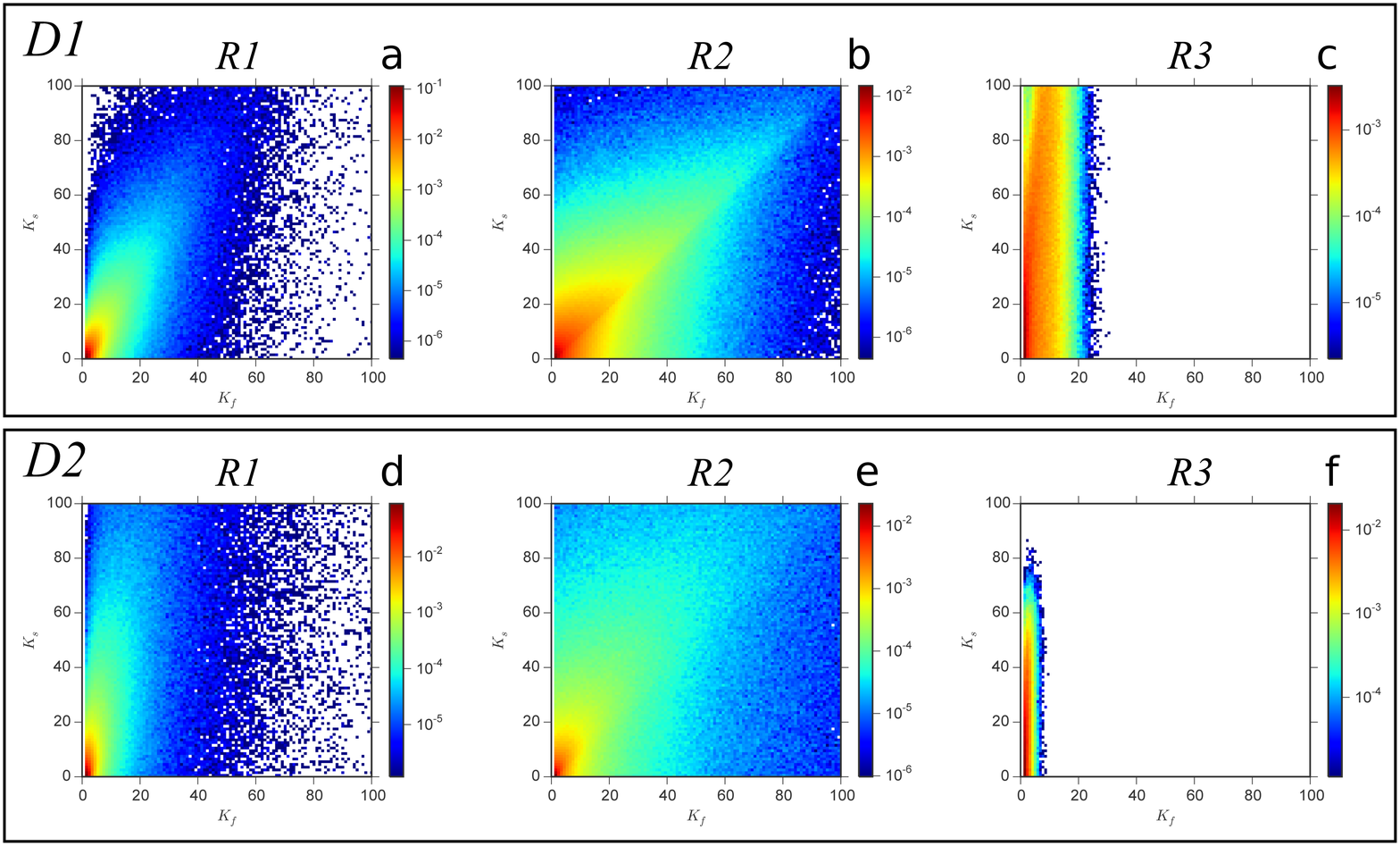}
\caption{\textbf{The rank-based analyses of randomized versions of the empirical datasets.} When we compare the ranks obtained from the $R1$ randomization of $D1$ and $D2$ \textbf{(a,d)}, we can observe that both distributions are very similar, sharing many common features whereas from the second randomization  \textbf{(b,e)} we can see that both $D1$ \textbf{(b)} and  $D2$ \textbf{(e)} have very different shapes in comparison with the empirical data as depicted in Figure \protect{\ref{fig:heatmaps}}. The data from the third randomization method \textbf{(c,f)} totally  deviates from the empirical data, as well as the other randomization results. It suggests that the patterns observed after the analyses of the visitation ranks distributions are indeed rooted on the way humans move.}
  \label{fig:random_heatmap_jet}
\end{figure*}

The first feature we can observe is that when we shuffle the trajectories in $D1$ (Figure \ref{fig:random_heatmap_jet}a), the ranks distribution exhibit a similar pattern as observed on the original data. However, it supports our claim that the predominance of the preferential return, as captured by the aggregated mobile phone data of $D1$, is hindering the micro-level dynamics characteristic of the recency effect. A closer look at the bottom rows of Figure \ref{fig:random_heatmap_jet}a does not show any increased probability due to recency. When we artificially destroy the power-law distribution of the visitation frequencies (Figure \ref{fig:random_heatmap_jet}b) we can observe a dramatic change in the ranks distribution. It suggests that a significant part of the ranks distribution of $D1$ is indeed rooted on the visitation frequencies, as predicted by the PR mechanism.

When we analyze the randomized versions of $D2$ the influence of the recency becomes even more evident. As before, shuffling the individuals trajectories (Figure \ref{fig:random_heatmap_jet}d) removes the features we described in Figure \ref{fig:heatmaps} (as before the evidence in the bottom rows are not there). Moreover, by removing the temporal information from visitation sequences in $D2$, the rank distributions acquire the same form as the one of $D1$.

\begin{figure*}[htb]
  \centering
  \includegraphics[width=0.9\linewidth]{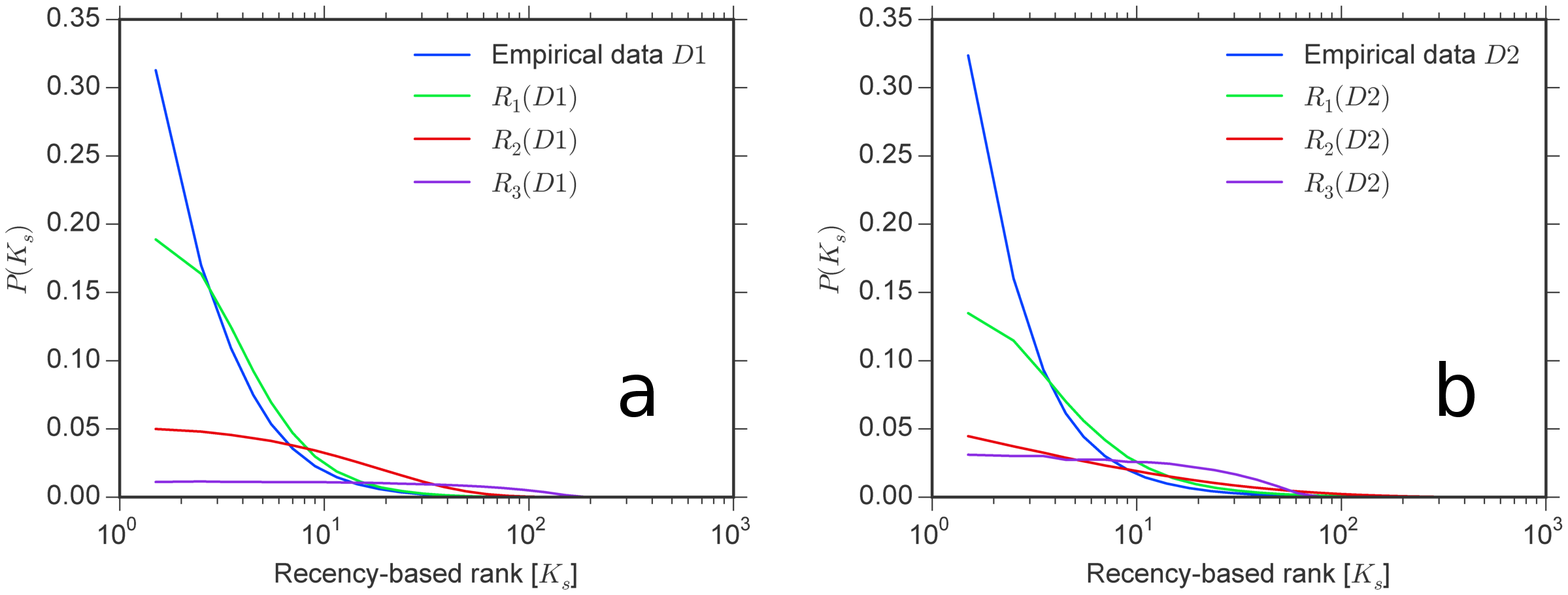}
  \caption{\textbf{Comparison of the distribution of the recency-based ranks generated after different randomizations of the original datasets $D1$ (a) and $D2$ (b).} The plots depict the PDF of the $K_{s}$ in log-linear scale.  We can observe that when we shuffle the individuals trajectories ($R1$), removing its temporal information---and hence any temporal effect such as recency---the upper part of the curves deviate from the empirical data.  When we remove the visitation frequency distribution ($R2$), the tail of the recency distribution also is destroyed, approximating to what we observe on the baseline curve ($R3$) where none of the original distributions is maintained.}
  \label{fig:recency_random_tests_d2}
\end{figure*}

When we look at the recency rank distributions for the randomized data (Figure \ref{fig:recency_random_tests_d2}), we see that the recency rank on the shuffled trajectories deviate from the empirical data, showing that the recency effect is indeed present in both datasets. More striking, however, is the fact that this analysis not only shows that the recency effect is limited to the most recently visited locations but also suggests a possible existence of an upper limit to the effect. More precisely, the recency effect could be stronger observed when returns occur after visiting 2 locations in $D1$ and 3 locations in $D2$. It means that if an individual is observed again in a recently discovered location, right after visiting less than 3 other locations, it is very likely that this location will become a frequently visited locations. 
  
In summary, our approach has shown strong evidence that returns in human trajectories are shaped by two distinct ingredients, one responsible for the long-term regularities (such as the PR mechanism) and another one to account for the changes in the visitation preferences, where recently visited locations also become highly visited location.
  

\subsection*{The Recency-based model}
To test to what extent the patterns we observed in the rank distribution corresponds to an unforeseen mechanism of human mobility, we tested for the hypothesis that it emerges from the data when we build the sequence-based ranks of frequency-driven trajectories. For the argument to hold true, the same patterns must be observed in the synthetic data produced by the EPR model. To test our hypothesis, we compared the purely frequentist mechanism of the EPR against our new human mobility model where returns have a bias toward recently-visited locations.

The recency-based model extends the preferencial return mechanism endowing it with a mechanism capable of capturing the visitation bias towards recently visited locations---all ingredients of the EPR model were kept intact except for the temporal dimension. The reason for that is because the waiting-time distribution of the EPR model determines only \emph{when} an individual is going to move (\ie, how much time he will wait still before the next jump) but not \emph{where} he goes. It is important to emphasize that the \emph{recency} bias underlying our model is regarding the visitation path and it is time-independent. 

The model can be described as follows: first, a population of $N$ agents is initialized and scattered randomly over a discrete lattice with $M\times M$ cells, each one representing a possible location. The initial position of each agent is accounted as its first visit. At each time step agents can either visit a new location if probability $p_{new} = \rho S^{\gamma}$ where $\rho = 0.6$ and $\gamma = 0.6$ are control parameters---whose values were derived by Song \et \cite{Song2010} from empirical data---and $S$ corresponds to the number of distinct locations visited thus far. With complementary probability $1 - p_{new}$ an agent return to a previously visited location. If the movement is selected to be a return, with probability $1 - \alpha$ the $i^\text{th}$ last visited location is selected from a Zipfian distribution (Zipf's law) with probability $p(i)\propto k_{s}(i)^{-\eta}$ where $k_{s}(i)$ is the recency-based rank of the location $i$. The parameter $\eta$ controls the number of previously visited locations a user would \emph{remember} when deciding to visit a location. With probability $\alpha$ the destination is selected based on the visitation frequencies with probability $\Pi_{i} \propto k_{f}(i)^{-1 -\gamma}$ where $k_{f}(i)$ is the frequency rank of location $i$. Notice that when $\alpha = 1$ we recover the original preferential return behavior of the EPR model while when $\alpha = 0$, visitation returns will be based solely on the recency. We experimentally tested different parameters configuration for the model. Our analyses have shown that when $\alpha = 0$, the heavy tail of  the visitation frequency disappears while for $\alpha = 1$ the power law of the recency  distribution vanishes. It suggests that both mechanisms must be present in order to reproduce those two features. In practice different individuals could have different $\alpha$ values.  However, extracting it from the empirical data is not an easy task once it is hard to determine either the movement was driven by the recency or frequency. Nevertheless, we determined that $\alpha = 0.1$ (\ie, 10\% of the movements influenced by the visitation frequencies) was enough to restore the recency and frequency ranks distributions. Also, for the Zipfian distribution of the recency rank we used $\eta = 1.6$, extracted from the empirical data. 

\begin{figure*}[htbp]
\centering
\includegraphics[width=0.9\linewidth]{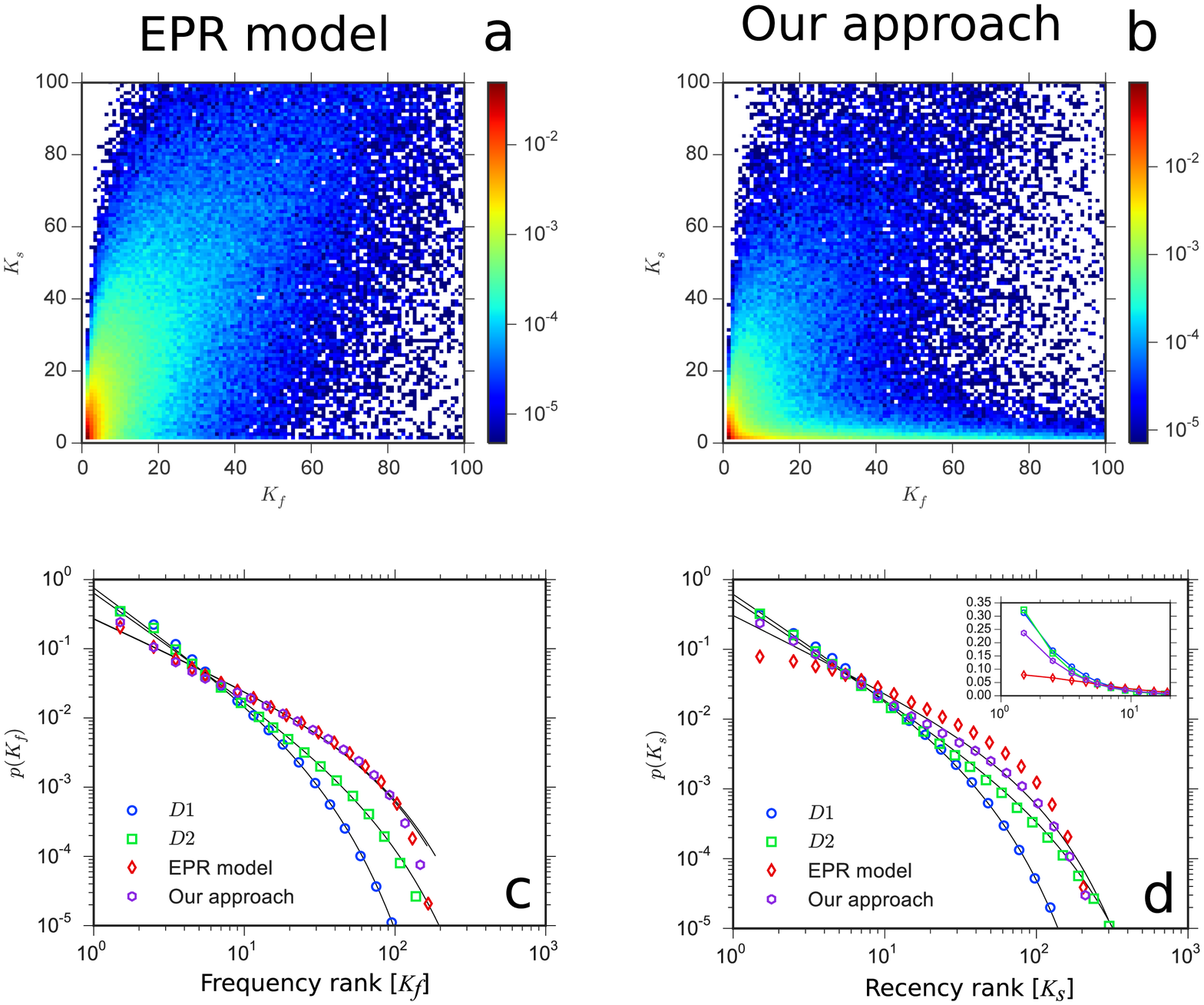}
\caption{\textbf{Comparison between the EPR model and the recency-based (RM) model}. \textbf{(a)} The analysis of the return ranks generated by the EPR model shows that it reproduces a pattern similar to the one observed from the empirical analysis, especially of $D1$. \textbf{(b)} On the other hand, on the presence of the recency mechanism, we can observe the same high probability of return to recently visited locations (\ie, low $K_{s}$) as observed on the empirical data. \textbf{(c)} When we look at the distribution of the frequency ranks, the preferencial return mechanisms (red diamonds) successfully exhibited a power-law distribution, in agreement with the empirical observations. The activation of the recency mechanism does not affect the frequency rank distribution (purple hexagons). \textbf{(d)} However, when we look at the $K_{s}$ distribution, the EPR mechanism does not capture the power-law behavior observed on the empirical data.}
\label{fig:order_order_d1_d2_d3_im_rm_order}
\end{figure*}

Visually, the synthetic data produced by the EPR model seems to have a good approximation with the empirical data (see Figure \ref{fig:order_order_d1_d2_d3_im_rm_order}). However, when we compare the bottom-most rows of the histogram, it deviates from the empirical evidence, by not capturing the broader distribution of $p(k_{f},k_{s})$ for recently visited locations. On the other hand, the recency-based mechanism (RM) reproduced the recency influence as observed in the empirical data (Figure \ref{fig:order_order_d1_d2_d3_im_rm_order} \textbf{b}). When we look at the $K_{f}$ distribution, the EPR model recovers its heavy tail, as one would expect  (Figure \ref{fig:order_order_d1_d2_d3_im_rm_order} \textbf{d}). On the other hand, when we look at each variable individually we notice that the $K_{s}$  distribution as produced by the EPR model deviates from a power law. In fact, it is better approximated by an exponential distribution whereas recency-model maintains its power-law behavior.  The differences in the $K_{s}$ distribution as produced by both models become more evident in log-linear scale (inset of Figure  \ref{fig:order_order_d1_d2_d3_im_rm_order} \textbf{d}), where we can clearly see that the EPR model does not capture the preference for  recently visited location.

\begin{figure}[htbp]
\centering
\includegraphics[width=0.8\linewidth]{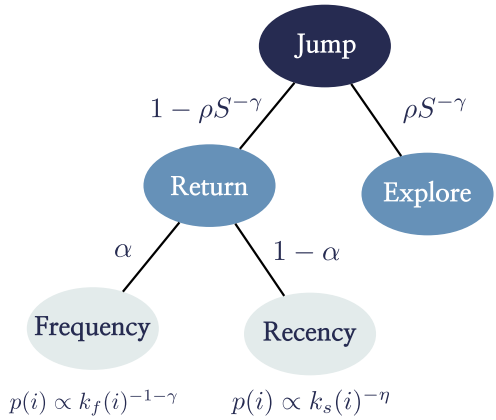}
\caption{\textbf{Recency-based individual mobility model.} Notice that the exploration mechanism is  kept the same as in the EPR model. In addition to the PR mechanism, the proposed model incorporates the recency effect, where recently-visited locations have also a high visitation probability.}
\end{figure}

\section*{Discussion}
When we look at an individual's trajectories over, let us say, one year, the visitation patterns and regularities become very evident and radical changes in visitation patterns---such as during a long vacation abroad or after starting a new job in another city---are very unlikely. In a large population, these events indeed occur, but their effect on the population scale are very diluted and, sometimes, transient. Within such limited time window, individuals indeed are predictable, and believing that one is going to be at one of its most visited locations is a reasonable guess. However, it is really unlikely that the individual's preferences  are the same for 10 or 20 years. A recently-discovered quality restaurant is a more plausible destination than our former workplace. Some events in our lives have the potential to reshape not only our visitation patterns but also our  preferences. In this work we explored this idea under a simple rank-based framework. We unveiled empirical evidences supporting the idea that human trajectories are biased towards recently visited locations. We also offered a different perspective for human mobility investigation, where the temporal dimension plays a role much more important than the inter-event times. 

\section*{Authors' contributions}
Developed the ideas, methods and analyses: HB and RM.  Empirical data analysis: HB and AE. Wrote the manuscript: HB, FBLN and RM.


\begin{thebibliography}{10}

\bibitem{Balcan2011}
Duygu Balcan and Alessandro Vespignani.
\newblock {Phase transitions in contagion processes mediated by recurrent
  mobility patterns}.
\newblock {\em Nature physics}, 7(March), 2011.

\bibitem{Barabasi1999}
Albert-L\'{a}szl\'{o} Barab\'{a}si and R\'{e}ka Albert.
\newblock {Emergence of scaling in random networks}.
\newblock {\em Science}, page~11, October 1999.

\bibitem{Belik2011}
Vitaly Belik, Theo Geisel, and Dirk Brockmann.
\newblock {Natural Human Mobility Patterns and Spatial Spread of Infectious
  Diseases}.
\newblock {\em Physical Review X}, 1(1):011001, August 2011.

\bibitem{Brockmann2006}
Dirk Brockmann, L~Hufnagel, and T~Geisel.
\newblock {The scaling laws of human travel.}
\newblock {\em Nature}, 439(7075):462--5, January 2006.

\bibitem{Cho2011}
Eunjoon Cho, Seth~A. Myers, and Jure Leskovec.
\newblock {Friendship and mobility}.
\newblock In {\em Proceedings of the 17th ACM SIGKDD international conference
  on Knowledge discovery and data mining - KDD '11}, page 1082, New York, New
  York, USA, 2011. ACM Press.

\bibitem{Clauset2009}
Aaron Clauset, Cosma~Rohilla Shalizi, and Mark E.~J. Newman.
\newblock {Power-Law Distributions in Empirical Data}.
\newblock {\em SIAM Review}, 51(4):661--703, November 2009.

\bibitem{Colizza2007}
Vittoria Colizza, Alain Barrat, Marc Barthelemy, Alain-Jacques Valleron, and
  Alessandro Vespignani.
\newblock {Modeling the worldwide spread of pandemic influenza: baseline case
  and containment interventions.}
\newblock {\em PLoS medicine}, 4(1):e13, January 2007.

\bibitem{Gonzalez2008a}
Marta~C. Gonz\'{a}lez, C\'{e}sar~A. Hidalgo, and Albert-L\'{a}szl\'{o}
  Barab\'{a}si.
\newblock {Understanding individual human mobility patterns}.
\newblock {\em Nature}, 453(7196):479--482, June 2008.

\bibitem{Grabowicz2013}
PA~Grabowicz, JJ~Ramasco, Bruno Gon\c{c}alves, and VM~Egu\'{\i}luz.
\newblock {Entangling mobility and interactions in social media}.
\newblock {\em PloS one}, pages 1--16, 2014.

\bibitem{Hasan2012a}
Samiul Hasan, Christian~M. Schneider, Satish~V. Ukkusuri, and Marta~C.
  Gonz\'{a}lez.
\newblock {Spatiotemporal Patterns of Urban Human Mobility}.
\newblock {\em Journal of Statistical Physics}, 151:304--318, 2012.

\bibitem{Jung2008}
WS~Jung, Fengzhong Wang, and HE~Stanley.
\newblock {Gravity model in the Korean highway}.
\newblock {\em EPL (Europhysics Letters)}, pages 1--13, 2008.

\bibitem{Kitamura2000}
Ryuichi Kitamura, Cynthia Chen, RM~Pendyala, and Ravi Narayanan.
\newblock {Micro-simulation of daily activity-travel patterns for travel demand
  forecasting}.
\newblock {\em Transportation}, pages 25--51, 2000.

\bibitem{Krajzewicz2011a}
Daniel Krajzewicz, Georg Hertkorn, Peter Wagner, and Christian R\"{o}ssel.
\newblock {SUMO ( Simulation of Urban MObility ) An open-source traffic
  simulation Car-Driver Model}.
\newblock 2011.

\bibitem{Krumme2013}
Coco Krumme, Alejandro Llorente, Manuel Cebrian, Alex~Sandy Pentland, and
  Esteban Moro.
\newblock {The predictability of consumer visitation patterns.}
\newblock {\em Scientific reports}, 3:1645, January 2013.

\bibitem{Lenormand2015}
Maxime Lenormand, Bruno Gon\c{c}alves, Ant\`{o}nia Tugores, and Jos\'{e}~J.
  Ramasco.
\newblock {Human diffusion and city influence}.
\newblock {\em arXiv}, pages 1--17, 2015.

\bibitem{Lu2013}
Xin Lu, Erik Wetter, Nita Bharti, Andy Andrew~J Tatem, and Linus Bengtsson.
\newblock {Approaching the limit of predictability in human mobility.}
\newblock {\em Scientific reports}, 3:2923, January 2013.

\bibitem{Merton1968}
R~K Merton.
\newblock {The Matthew Effect in Science}.
\newblock {\em Science (New York, N.Y.)}, 159:56--63, 1968.

\bibitem{Price1976}
DS~Price.
\newblock {A general theory of bibliometrics and other cumulative advantage
  processes}.
\newblock {\em Journal of the American Society for Information \ldots}, 1976.

\bibitem{Sadilek2012}
Adam Sadilek and John Krumm.
\newblock {Far Out: Predicting Long-Term Human Mobility.}
\newblock {\em AAAI}, pages 814--820, 2012.

\bibitem{Schneider2013}
Christian~M. Schneider, Vitaly Belik, Thomas Couronn\'{e}, Zbigniew Smoreda,
  Marta~C. Gonz\'{a}lez, and Thomas Couronne.
\newblock {Unravelling daily human mobility motifs.}
\newblock {\em Journal of the Royal Society, Interface / the Royal Society},
  10:20130246, 2013.

\bibitem{Song2010}
Chaoming Song, Tal Koren, Pu~Wang, and Albert-l\'{a}szl\'{o} Barab\'{a}si.
\newblock {Modelling the scaling properties of human mobility}.
\newblock {\em Nature Physics}, 6(10):818--823, September 2010.

\bibitem{Song2010b}
Chaoming Song, Zehui Qu, Nicholas Blumm, and Albert-L\'{a}szl\'{o}
  Barab\'{a}si.
\newblock {Limits of predictability in human mobility.}
\newblock {\em Science (New York, N.Y.)}, 327(5968):1018--21, February 2010.

\bibitem{Szell2012}
Michael Szell, Roberta Sinatra, Giovanni Petri, Stefan Thurner, and Vito
  Latora.
\newblock {Understanding mobility in a social petri dish}.
\newblock {\em Scientific Reports}, 2:1--6, 2012.

\bibitem{Toole2012}
Jameson~L. Toole, M.~Ulm, Marta~C. Gonz\'{a}lez, and D.~Bauer.
\newblock {Inferring land use from mobile phone activity}.
\newblock {\em Proceedings of the ACM SIGKDD international workshop on urban
  computing}, pages 1--8, 2012.

\bibitem{Wang2011}
Dashun Wang, Dino Pedreschi, Chaoming Song, Fosca Giannotti, and
  Albert-L\'{a}szl\'{o} Barab\'{a}si.
\newblock {Human mobility, social ties, and link prediction}.
\newblock In {\em Proceedings of the 17th ACM SIGKDD international conference
  on Knowledge discovery and data mining - KDD '11}, page 1100, New York, New
  York, USA, 2011. ACM Press.
  
\bibitem{Hsu2006}
Wei~Jen Hsu and Ahmed Helmy.
\newblock {On modeling user associations in wireless lan traces on university
  campuses}.
\newblock {\em 2006 4th International Symposium on Modeling and Optimization in
  Mobile, Ad Hoc and Wireless Networks, WiOpt 2006}, (June 2015), 2006.

\bibitem{Yang2014}
Yingxiang Yang, Carlos Herrera, Nathan Eagle, and Marta~C. Gonz\'{a}lez.
\newblock {Limits of predictability in commuting flows in the absence of data
  for calibration.}
\newblock {\em Scientific reports}, 4:5662, 2014.

\end{thebibliography}

\begin{thebibliography}{1}

\bibitem{Clauset2009}
Aaron Clauset, Cosma~Rohilla Shalizi, and Mark E.~J. Newman.
\newblock {Power-Law Distributions in Empirical Data}.
\newblock {\em SIAM Review}, 51(4):661--703, November 2009.

\bibitem{Mukund2008a}
Seshadri Mukund, Sridhar Machiraju, Ashwin Sridharan, Jean Bolot, Christos
  Faloutsos, and Jure Leskovec.
\newblock {Mobile Call Graphs: Beyond Power-Law and Lognormal Distributions}.
\newblock 2008.

\bibitem{Reed2004}
William~J. Reed and Murray Jorgensen.
\newblock {The Double Pareto-Lognormal Distribution—A New Parametric Model
  for Size Distributions}.
\newblock {\em Communications in Statistics - Theory and Methods},
  33(8):1733--1753, 2004.

\bibitem{smirnov1948}
N.~Smirnov.
\newblock Table for estimating the goodness of fit of empirical distributions.
\newblock {\em Ann. Math. Statist.}, 19(2):279--281, 06 1948.

\bibitem{smirnov1939estimation}
Nikolai~V Smirnov.
\newblock On the estimation of the discrepancy between empirical curves of
  distribution for two independent samples.
\newblock {\em Bull. Math. Univ. Moscou}, 2(2), 1939.

\end{thebibliography}

\clearpage

\appendix
\appendixpage
\renewcommand\thefigure{\thesection\arabic{figure}}   
\renewcommand\thetable{\thesection\arabic{table}}   
\setcounter{figure}{0}
\setcounter{table}{0}
\section{Statistical analysis of the rank distributions}
\label{appendix:stats}
As presented in the main text, the probability distributions of the rank variables $K_{f}$ and $K_{s}$ are very similar. In order to assess whether the two rank variables come from the same  distribution, we performed the two-sided Kolmogorov-Smirnov test. To test our hypothesis (i.e., both variables come from the same distribution), we compare the Kolmogorov-Smirnov distance $D_{K_{f},K_{s}}$ with the critical value $D_{\alpha}$ for a desired $\alpha$ level where $$D_{\alpha} = c(\alpha) \sqrt{\frac{n_{1}+n_{2}}{n_{1}n_{2}}},$$ with sample size $n_{1} = n_{2} = n$ and $c(\alpha) = 1.95$ for $\alpha = 0.001$ \cite{smirnov1948,smirnov1939estimation}. As one can see from the Table \ref{tab:rank_test}, the distance  $D_{K_{f},K_{s}} > D_{\alpha}$ for both datasets, rejecting the hypothesis that $K_{f}$ and $K_{s}$ were drawn from the same distribution.

\begin{table}[htp]
  \centering
  \caption{Two-sided Kolmogorov-Smirnov test and $p$-values for the rank variables.}\label{tab:rank_test}
  \begin{tabular}{@{}lcccc@{}}
    \toprule
    Dataset                 & $n$ &  $D_{K_{f},K_{s}}$ & $D_{\alpha}$    & $p$-value \\
    \midrule
    $D1$ & 564228   & 0.07999& 0.00367 & 0.0       \\
    $D2$ &  2267116   &  0.09708    &0.00183& 0.0        \\
    \bottomrule                                   
  \end{tabular}
\end{table}

Additionally, in order to determine the probability function that better characterizes the rank variables, we compared the goodness of fit offered by different heavy-tailed distributions. 
For this test, we measured the fit provided by two other distributions, namely log-normal and double-Pareto log-normal (dPlN). In some scenarios, the log-normal and the truncated power law distributions can both yield very similar results. 

More recently the dPlN distribution has been reported to offer a very sound model for many empirical data such as income distribution, oil-field sizes \cite{Reed2004} and the degree distribution of social networks \cite{Mukund2008a}. The double-Pareto log-normal corresponds to a mixture of two power laws joined by a log-normal segment \cite{Reed2004}. The PDF of the dPlN can be defined as 
\begin{equation}
f(x) = \frac{\alpha\beta}{\alpha+\beta} \left[f_{1}(x) + f_{2}(x) \right],
\end{equation}
$$\quad  f_{1}(x) = x^{-\alpha - 1}e^{\alpha\nu + \alpha^{2}\tau^{2}/2} \Phi\left(\frac{\ln x - \nu - \alpha\tau^{2}}{\tau}\right),
 $$
 
 $$
 \quad f_{2}(x) = x^{\beta -1}e^{-\beta\nu + \beta^{2}\tau^{2}/2 }\Phi^{c}\left(\frac{\ln x - \nu + \beta \tau^{2}}{\tau}\right), $$
where $\Phi$ is the CDF (Cumulative Distribution Function) of the standard normal $N(0,1)$ and $\Phi^{c}$ is the CCDF of $N(0,1)$.

\begin{figure}[htbp]
\centering
\includegraphics[width=0.45\linewidth]{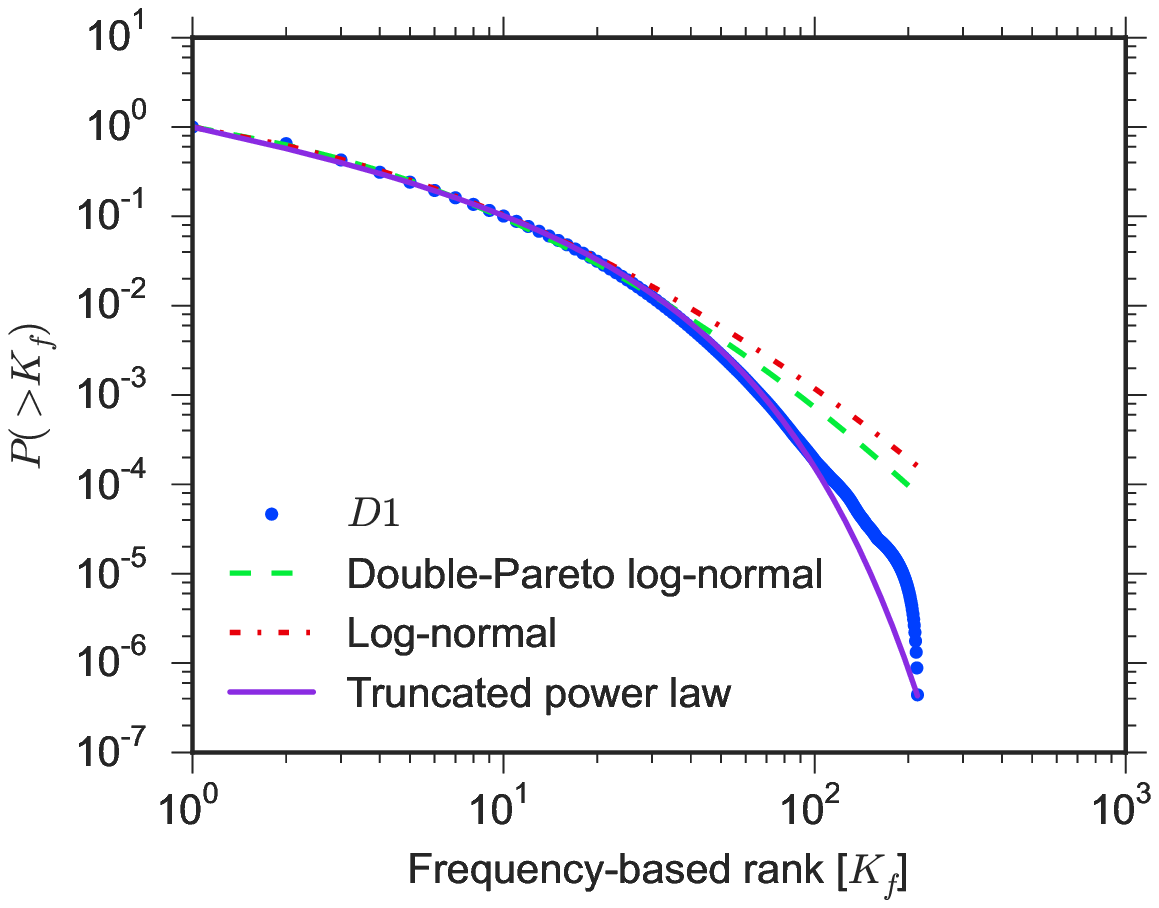}
\includegraphics[width=0.45\linewidth]{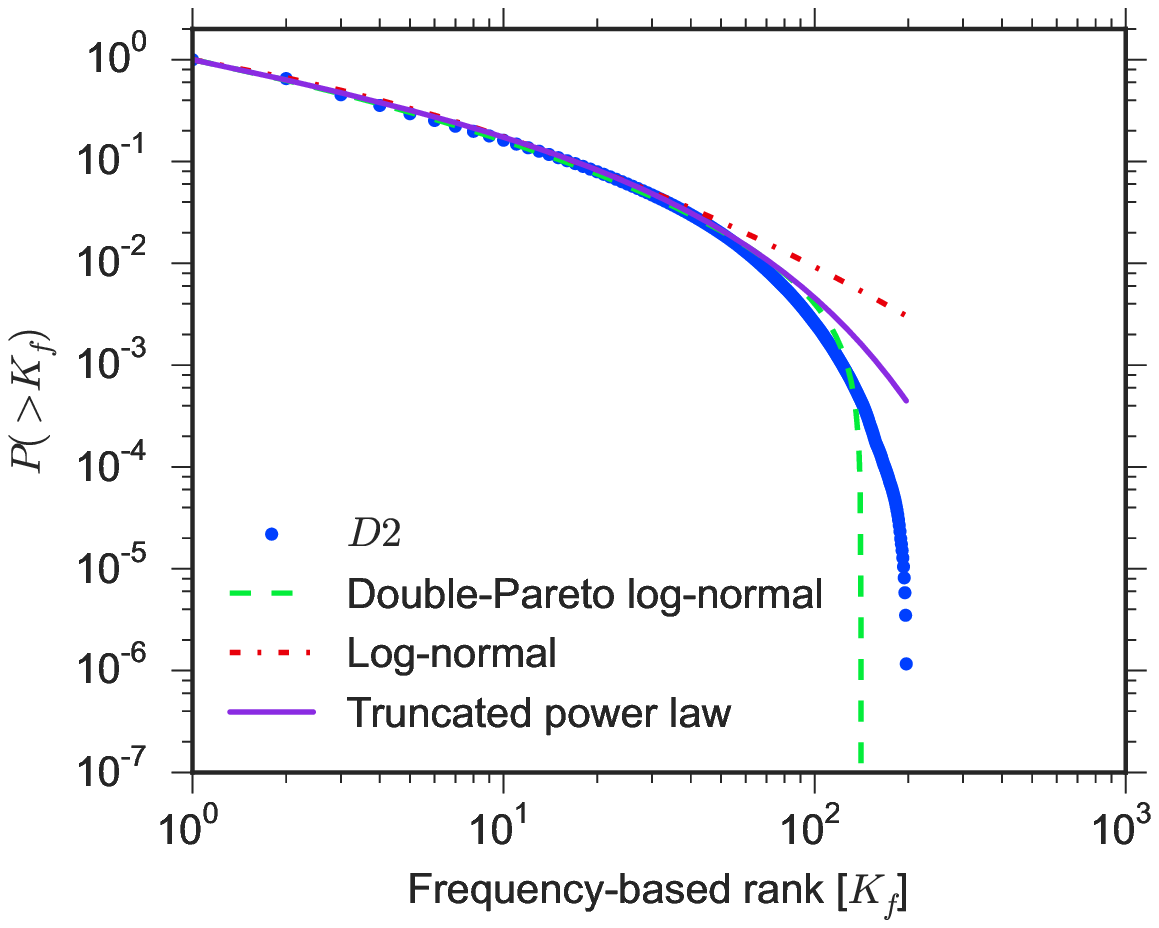}
\includegraphics[width=0.45\linewidth]{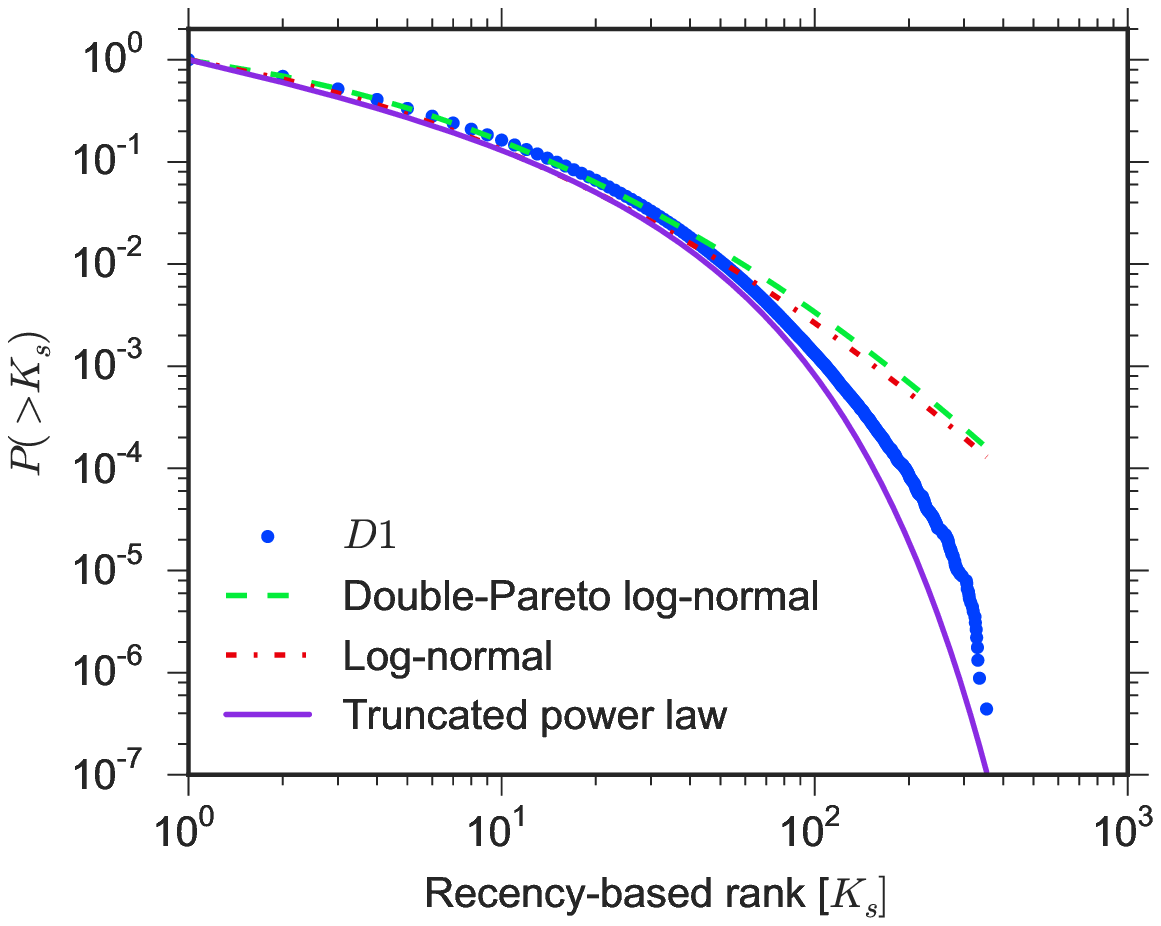}
\includegraphics[width=0.45\linewidth]{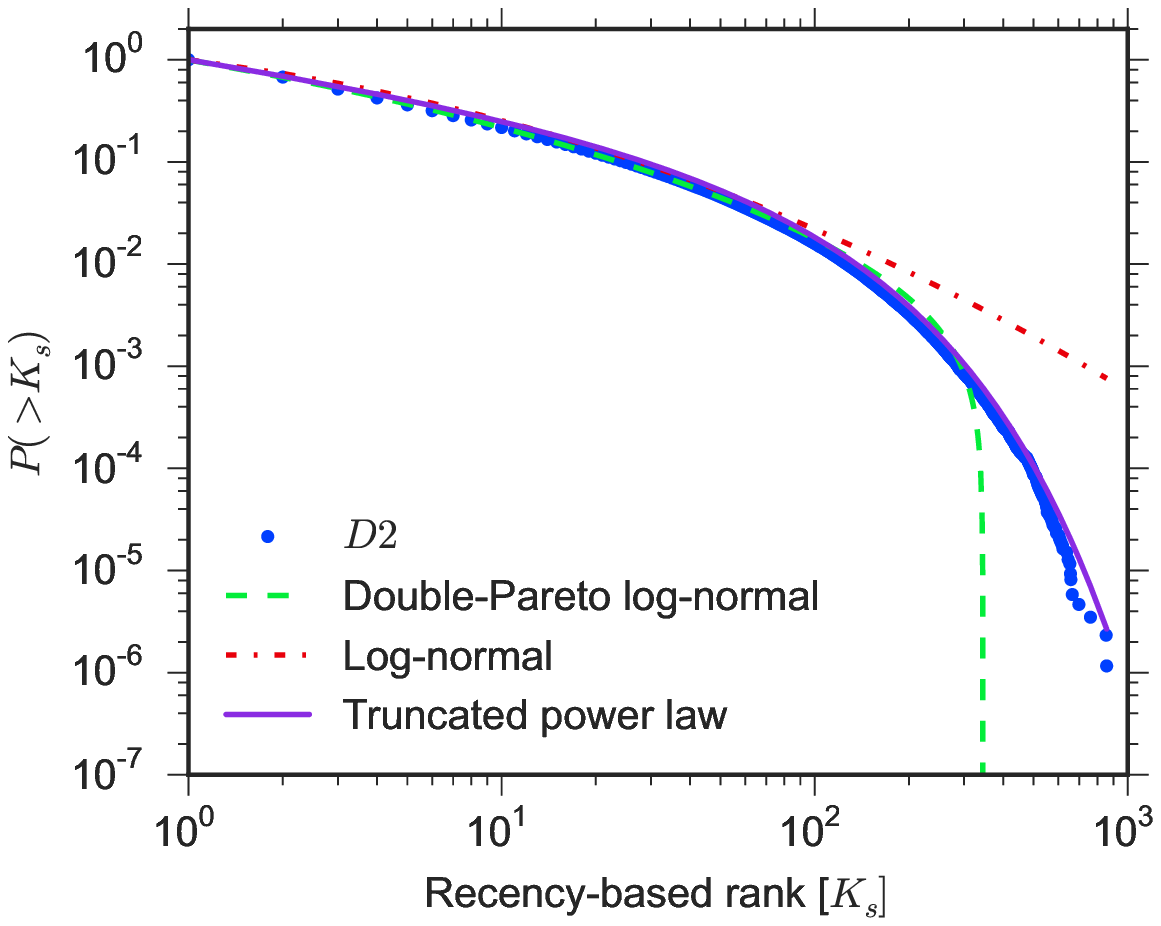}
\caption{Curve fits of different heavy-tailed distributions for both $K_{f}$ (top charts) and $K_{s}$ (bottom charts). In addition to the well-known log-normal (dot-dashed line) and truncated power law (solid line)  distributions, we also measured the goodness of fit for the double-Pareto log-normal.}
\label{fig:distribution_comparison_kf_d1}
\end{figure}

The log-likelihood ratio test compares two competing candidate distributions where the one with the higher likelihood is the one that provides the better fit. The sign of the log-likelihood ratio indicates the prevalence of the target distribution (here, the truncated power law) over an alternative competing hypotheses whereas the $p$-value indicates the significance level of the test \cite{Clauset2009}. As shown on the Table \ref{tab:goodness_ranks}, both rank variables are indeed better approximated by truncated power laws. 

\begin{table*}[ht]
\centering
\caption{Comparison between the goodness of fit provided by the truncated power law and other distributions via  log-likelihood ratio test.}
\label{tab:goodness_ranks}

\begin{tabular}{lllcc}
\toprule
Dataset & Rank& Alternative distribution& Log-likelihood ratio & $p$-value\\
\midrule
\multirow{4}{*}{$D1$} & \multirow{2}{*}{frequency}& Log-normal& 50.195& 0.0\\
& & Double-Pareto Log-normal& 157.185& 0.0\\
 & \multirow{2}{*}{recency}& Log-normal& 46.147& 0.0\\
 & & Double-Pareto log-normal &68.077& 0.0\\
 \midrule
\multirow{4}{*}{$D2$}&  \multirow{2}{*}{frequency}& Log-normal&114.455 & 0.0\\
& & Double-Pareto log-normal& 90.703 & 0.0\\
 & \multirow{2}{*}{recency}& Log-normal& 45.58& 0.0\\
 & & Double-Pareto log-normal & 128.884& 0.0\\
\midrule
\end{tabular}
\end{table*}


\section{Parameters estimation}
\label{appendix:parameters}
The probability distribution of the rank-based variables described in the main text were better approximated by truncated power-law distributions  $p(x) = Cx^{-\alpha}\mathrm{e}^{-x/\tau}$. Parameters were estimated using the methods in Ref. \cite{Clauset2009}. 

\begin{table}[htp]
\centering
\caption{Estimated parameters of the truncated power-law distributions with the best fit for the rank variables.}\label{tab:rank_fits}
\begin{tabular}{@{}llcc@{}}
\toprule
Dataset               & Rank    & $\alpha$ & $\tau$ \\
\midrule
\multirow{2}{*}{$D1$} & recency & 1.644    & 41.66  \\
                      & frequency& 1.859    & 37.0         \\\midrule
\multirow{2}{*}{$D2$} & recency &  1.699   &  250.0         \\
                      & frequency&  1.625   & 125.0 \\ 
                      \bottomrule                                   
\end{tabular}
\end{table}

\end{document}